\RequirePackage{fix-cm}
\documentclass[smallextended]{svjour3}

\smartqed

\usepackage{graphicx}
\usepackage[margin=2.5cm, centering]{geometry}
\usepackage{mathptmx}
\usepackage{float}
\usepackage{caption}
\usepackage[english]{babel}
\usepackage{csquotes}
\usepackage[compress]{cite}
\usepackage{rotating}
\usepackage{afterpage}

\begin{document}

\title{Optimum optical designs for diffraction-limited terahertz spectroscopy and imaging systems using off-axis parabolic mirrors}

\author{Nishtha Chopra \and
	James Lloyd-Hughes 
}

\institute{N.\,Chopra \at
	Department of Physics, University of Warwick, Coventry, United Kingdom
	\email{nishtha.chopra@warwick.ac.uk}
	\and
	J.\,Lloyd-Hughes \at
	Department of Physics, University of Warwick, Coventry, United Kingdom 
    \email{j.lloyd-hughes@warwick.ac.uk}
 }

\date{Received: date / Accepted: date}

\maketitle

\begin{abstract}
Off-axis parabolic mirrors (OAPMs) are widely used in the THz and mm-wave communities for spectroscopy and imaging applications, as a result of their broadband, low-loss operation and high numerical apertures. 
However, the aspherical shape of an OAPM creates significant geometric aberrations that make achieving diffraction-limited performance a challenge, and which lowers the peak electric field strength in the focal plane.
Here we quantify the impact of geometric aberrations on the performance of the most widely-used spectrometer designs, by using ray tracing and physical optics calculations to investigate whether diffraction-limited performance can be achieved in both the sample and the detector plane.
We identify simple rules, based on marginal ray propagation, that allow spectrometers to be designed that are more robust to misalignment errors, and which have minimal aberrations for THz beams.
For a given source this allows the design of optical paths that give the smallest THz beam focal spot, with the highest THz electric field strength possible. 
This is desirable for improved THz imaging, for better signal-to-noise ratios in linear THz spectroscopy and optical-pump THz-probe spectroscopy, and to achieve higher electric field strengths in non-linear THz spectroscopy.

\end{abstract}

\section{Introduction}

Within the research community, different optical setups containing off-axis parabolic mirrors (OAPMs) are widely utilized to characterise broadband THz radiation sources \cite{Mosley2017,Rouzegar2023,chopra}, as well as to perform linear THz spectroscopy in the transmission or reflection geometry \cite{lee2009principles,Sun2018}, non-linear THz spectroscopy\cite{Reimann2021}, optical-pump THz-probe spectroscopy (OPTPS) \cite{Ulbricht2011} and near-field THz microscopy \cite{Lloyd-Hughes2021}.
The optimum performance for a THz spectrometer will be achieved if an aberration-free image of the THz source can be formed at each focal plane. 
For linear THz spectroscopy, a minimal spot size is often desired: to investigate samples that are small in transverse extent \cite{Jones2014}; to couple more effectively to sub-wavelength structures and waveguides\cite{Lloyd-Hughes2009}; or to image spatially inhomogeneous materials, such as large-area graphene \cite{Tomaino2011}. 
Further, the signal detected in the detector plane (\emph{e.g.}\ via electro-optic sampling or photoconductive detection) in THz time-domain spectroscopy is proportional to the electric field of the THz beam (rather than the area- and time-integrated power, as in a bolometer or pyroelectric), and is hence larger when the THz beam has a smaller area.
In OPTPS, the THz probe beam must have a smaller spatial extent than the optical pump beam, in order to probe a uniform carrier density: it is therefore also desirable to have a smaller THz beam so this condition can be readily achieved for lower power optical pump lasers \cite{Butler-Caddle2023}. 
Finally, intense THz pulses with high electric field strengths (typically $>100$\,kV/cm) can be used to study the non-linear dynamical motion of vibrational modes or free charges \cite{Reimann2021}.
For non-linear THz spectroscopy, it is thus important to efficiently focus one or more THz beams down to as close to the diffraction limit as possible, to obtain the highest electric field strength.

The majority of sources of broadband THz radiation (such as photoconductive antennae, optical rectification in non-linear crystals and spintronic emitters) produce THz beams that propagate in free space in the fundamental (TEM$_{00}$) transverse Gaussian mode. The Gaussian nature of terahertz (THz) beams has been corroborated through a combination of theoretical and experimental techniques, wherein, scalar and vectorial diffraction theory has been utilized to analyze the far-field behaviour of the THz pulses \cite{Gurtler:00}. Similarly, experimental validation was achieved by imaging the spatial and temporal distribution of pulses, in a standard THz-TDS setup \cite{Jiang:99}.  
In free space, the divergence angle of this Gaussian mode is $\theta = \lambda / \pi w_0$ in the paraxial approximation, where $w_0$ is the radius of the initial beam waist. 
For example, $\theta=18^{\circ}$ for light at 1\,THz ($\lambda=300$\,$\mu$m) for $w_0=300$\,$\mu$m, and the divergence angle increases for longer wavelengths or smaller initial beam sizes.
Hence, THz radiation tends to diverge rapidly, and large numerical aperture optics are required, in particular, to collect low-frequency components with longer wavelengths \cite{peiponen2012terahertz}. 
While polymer lenses are widely adopted in many THz systems due to their light weight and ease of fabrication \cite{lee2009principles}, their finite Fresnel losses and absorption reduces optical throughput, and they can introduce optical aberrations. 

\begin{figure}[tb]
	\centering
	\includegraphics[width=0.55\textwidth]{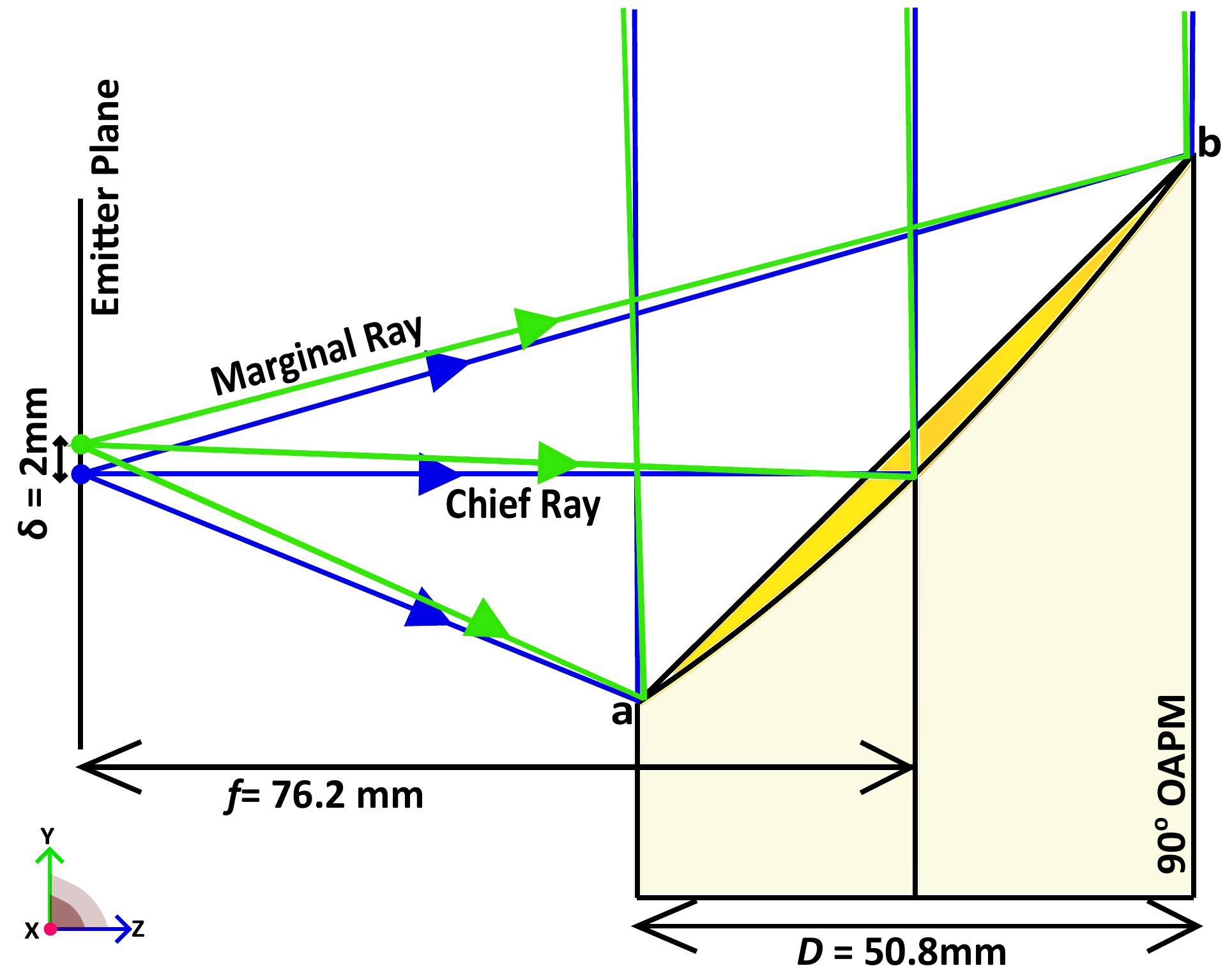}
	\caption{Geometry of a right-angle OAPM.}
	\label{f:OAPM}       
\end{figure}

To overcome these challenges, high-reflectivity metal-coated off-axis parabolic mirrors (OAPMs) are widely used \cite{peiponen2012terahertz}. 
An OAPM is a segment of a parent parabolic surface and can be defined by the off-axis angle, reflected effective focal length $f$ and diameter $D$, as illustrated in Fig.\ \ref{f:OAPM}.
These mirrors offer broadband performance while eliminating Fresnel losses, and can collimate a beam diverging from an on-axis point source without aberrations, or focus a collimated beam to a diffraction-limited spot.
In the geometric optics (ray optics) picture, an OAPM can perfectly collimate rays of light that originate from a point source only if the point source is on the optical axis and in the focal plane, at a distance $f$ from the mirror (blue rays in Fig.\ \ref{f:OAPM}).
However, real THz sources have a finite spatial extent, set by the size of the incident laser beam or the emitter's active area in laser-based THz systems, and the propagation of light from off-axis points in the object plane must also be considered (green rays in Fig.\ \ref{f:OAPM}).

Different optical analysis techniques have been deployed to model OAPM systems, such as deriving analytical expressions for Hermite-Gaussian beam propagation after reflection from an OAPM in the paraxial approximation \cite{Murphy1987} or by using optical ray tracing methods \cite{Arguijo2003,Bruckner2010}.
In the former case, after a fundamental Gaussian beam was incident onto an OAPM, the reflected electric field was found to have a skewed and broadened profile corresponding to the presence of additional higher order (TEM$_{30}$ and TEM$_{12}$) Hermite-Gaussian modes \cite{Murphy1987}. 
Importantly, by adding a second OAPM placed at the correct distance and with the right orientation, it was shown that the distortions introduced by the second mirror can cancel out the distortions introduced by the first mirror, resulting again in a clean fundamental Gaussian beam (desirable in order to achieve a diffraction-limited spot).
Alternatively, analytical ray tracing was used to show that if a collimated beam of light is incident onto an OAPM at an angle away from its optical axis, the focal spot is distorted \cite{Arguijo2003}. 

While analytical approaches are useful for specific cases where the underlying assumptions are valid, they become cumbersome for realistic optical systems containing many OAPMs. 
An alternative approach is to use optics modelling software to simulate THz beam propagation, such as performed using ray-tracing to model OAPMs \cite{Bruckner2010}, or using point spread functions to simulate bi-conic curved mirrors that produce a line focus \cite{Jordens2006}. 
In modelling packages such as ZEMAX, the optical performance of beams with finite spatial extent can be modelled using ray tracing and point sources that are placed off-axis, or via physical optics propagation, which takes coherent propagation effects (diffraction and interference) into account. 
Br\"{u}ckner \emph{et al.} used the ray tracing approach in the ZEMAX software to show that for an off-axis point source, the beam after a single right-angled OAPM has an astigmatic wavefront \cite{Bruckner2010}. 
Further, and similar to the conclusion derived in Murphy's work using analytical theory \cite{Murphy1987}, Br\"{u}ckner \emph{et al.} demonstrated that a judiciously-oriented second OAPM can cancel out the astigmatic wavefront, leading to diffraction-limited performance for off-axis field points. 
It was shown that the second OAPM, which was placed in a 4f geometry, has to be oriented to send the reflected beam anti-parallel to the original beam direction (i.e.\ along $-z$ if the THz source radiates along $z$) in order to substantially cancel out aberrations for off-axis field points.
This was in contrast to the alternative orientation for the second OAPM, reflecting the THz beam along $+z$, in which aberrations were evident.

Based on these considerations, Laurita \emph{et al.} subsequently performed an experimental comparison of the two geometries and showed that the beam waist at the sample focus of a typical THz-TDS system was smaller, at around 7\,mm compared to 11\,mm, for the aberration-corrected arrangement than the alternative orientation \cite{laurita2016modified}.
However, diffraction-limited performance was not achieved and the performance of the different OAPM arrangements in the detector plane was not considered. Recently, the present authors experimentally investigated a linear array of photoconductive THz emitters, where each pixel acted as a THz source at a different distance from the optical axis \cite{chopra}, and reported close to diffraction-limited performance obtained using an arrangement of OAPMs that corrected for aberrations. 

Given the above challenges for optical systems containing OAPMs, it is timely and pertinent to consider how best to minimise geometric aberrations, such that diffraction-limited performance can be achieved at both the sample position and the detector position in THz spectrometers.
In this article, we provide a comprehensive analysis of how the design of a spectrometer impacts its performance, with particular emphasis on the spatial and temporal spread in the sample and detector focal planes. 
In Section 2 we describe the modelling approach taken: we used the ZEMAX optical design software to model beam propagation (using ray tracing and physical optics) for the two optical systems most commonly used in THz spectroscopy, as well as an uncommon but optimum design, starting from a variety of on- and off-axis points.
Further, we introduce a notation that captures the relative orientation of different OAPMs, based on marginal ray propagation.
We then, in Section 3, examine and evaluate the performance of the different spectrometer designs by using spatial plots of beam propagation, and optical path differences in the time-domain. 
In Section 4 we conclude our analysis by ranking the three competing optical designs in terms of their robustness to misalignment, and their ability to form accurate images of the THz source in the sample plane and the detector plane. 
Finally, we provide rules based on marginal ray propagation to allow the optimum arrangement of OAPMs to be deduced without the need for optical modelling, which we check for a two OAPM system including a reflective mirror.

\section{Modelling approach and spectrometer designs}

In this Section, we define the ray-tracing and physical optics methodologies used and describe the key nomenclature used in the optical design field. 
Illustrated in Fig.\ \ref{f:OAPM} is a right-angled mirror viewed in the $y-z$ plane, also termed the tangential plane.
The $x-z$ plane is referred to as the sagittal plane.
The ray that propagates from the focus in the emitter plane along $z$, along the optical axis of the OAPM, is referred to as the \emph{chief ray}: it changes direction by 90$^{\circ}$ after hitting the geometric centre of the OAPM, to travel along $y$.
\emph{Marginal rays} are incident on the edges of the OAPM, and span the maximum aperture as seen from the geometric centre.
The asymmetric shape of the OAPM is particularly evident when considering the two marginal rays that hit the closest side (at point $a$) or the furthest side (point $b$) of the OAPM, as the two marginal rays travel at different angles to the chief ray.
Here, all OAPMs were assumed to be right-angled, with $f=76.2$\,mm and $D=50.8$\,mm, representative of the mirrors most often used.
For this choice, the numerical aperture was 0.31, and the angle of the marginal rays to the chief ray was $\alpha \simeq \tan^{-1}(D/2f)=18.4^{\circ}$.
This is similar to the divergence calculated above for a 1\,THz Gaussian beam, suggesting that the majority of the THz radiation produced by the emitter would be collected by this OAPM, and that the OAPM would be filled effectively such that smaller diffraction-limited beam focii can be achieved.

Typical THz spectroscopy and imaging systems require at least two pairs of two OAPMs: the first pair collects the THz radiation from the source, and focuses it onto a sample, while the second pair collects the transmitted or reflected THz beam and focuses it onto a detector.
There are a substantial number of degrees of freedom possible for a complete spectrometer design, as each individual OAPM has six positional degrees of freedom (three translational and three rotational).
Here we examined the most common spectrometer designs, which consist of four right-angled OAPMs, with their optical axes all within the same plane (here denoted $y-z$). 
For each OAPM added to an optical setup there are two options for its orientation that keep the optical axis in the same plane: these change the beam direction by $\pm90^{\circ}$ relative to its previous direction in the $y-z$ plane.

\begin{figure}[!b]
	\centering
	\includegraphics[width=\textwidth]{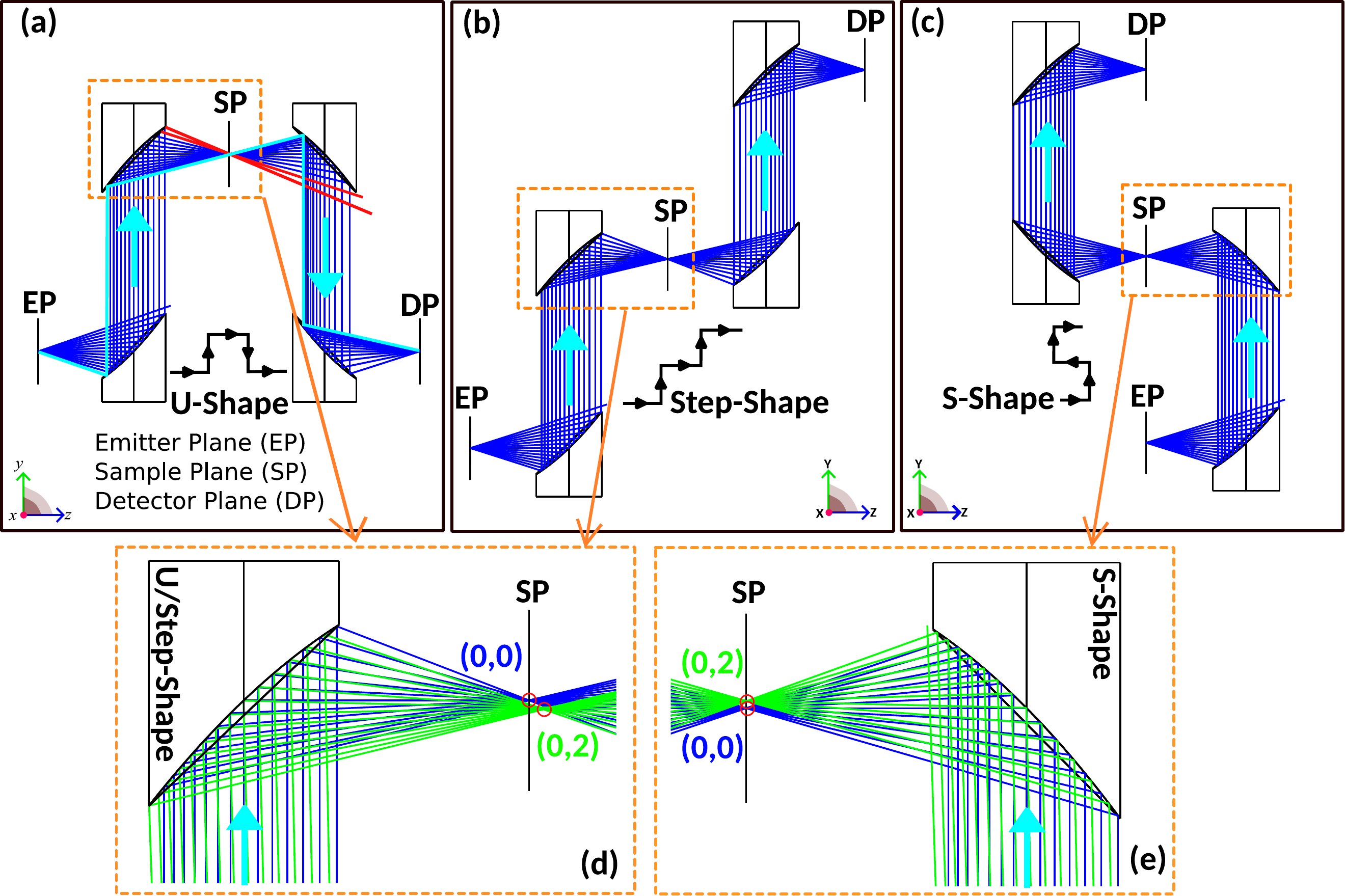}
	\caption{Ray tracing models of three different THz-TDS geometries for an on-axis point source at centred at $(0,0)$ in the emitter plane (EP), propagating to the sample plane (SP) and detector plane (DP). 
 (a) U-geometry, or $(a,b)(a,b)$ using the notation defined in the text. The red highlighted marginal rays are not captured as they propagate. 
 (b) Step-geometry, or $(a,b)(b,a)$ arrangement. The second OAPM pair is oriented the same as the first.
 (c) S-geometry, or $(a,a)(a,a)$ setup.
 (d) For the U- and step-shapes, propagation from an off-axis $(0,2$\,mm$)$ source, shown as green rays, leads to a tangential plane focus behind the SP focus for on-axis propagation, thus creating aberrations.
 (e) For the sample plane of the S-geometry, the off-axis rays rays from $(0,2$\,mm$)$ converge in the SP with minimal aberration.}
	\label{f:raytracing}       
\end{figure}

To describe three different geometries reported in the literature we define the nomenclature U-shape, step-shape and S-shape to refer to the spectrometers illustrated in Figs.\ \ref{f:raytracing}(a)-(c), based on the pattern formed when considering the chief ray's beam path (yellow lines and arrows).
The U-shape and step-shape design are the most widely used in THz time-domain spectrometers, while the S-shape was suggested to minimise aberrations in the sample plane \cite{laurita2016modified}.
Laurita \emph{et al.} compared the performance of the U-shape and S-shape (referred to as the ``conventional'' and ``modified'' designs in their work), but did not consider the step-shape or the performance in the detector plane.
In the U-shape and step-shape designs [Figs.\ \ref{f:raytracing}(a)-(b)], the first two OAPMs are oriented identically, but the designs differ in how the second pair of OAPMs is oriented. 
For the U-shape, the second pair of OAPMs is oriented to return the THz beam to the same $y$ co-ordinate, leading to a relatively compact design, while the step- and S-shapes have a larger size in $y$.

We stress here that the U-shape does not have the correct layout to collect all the \emph{on-axis} rays after the second OAPM: across the sample plane, the marginal rays highlighted in red are \emph{not} collected by the third OAPM [Figs.\ \ref{f:raytracing}(a)], and hence the throughput of the system is not perfect.
This problem is not seen for the step-shape or S-shape, where the third OAPMs are oriented correctly, and all the marginal rays after the sample plane are collected.

Based on a consideration of the marginal rays and the symmetry properties of an OAPM system, we define here a nomenclature that we find useful to describe the optical arrangement of pairs of OAPMs, inspired by the Glazer and Aleksandrov notation schemes widely used to describe the tilt patterns of perovskite octahedra.
As we show in the following section, the orientation of each OAPM within a pair and the relative orientation of each pair of OAPMs are both critical to achieve minimal aberrations.
Therefore, we introduce the notation $(x_i,x_j)$ to denote the orientation of a pair of OAPMs, where $x=a,b$ to denote whether or not the marginal ray reflects from the ``near'' side or ``far'' side points $a$ and $b$ in Fig.\ \ref{f:OAPM}, and the subscript denotes the focal lengths $f_i$ and $f_j$. 
If the focal lengths of all mirrors are identical, as in this work, we drop the subscript notation.
With this scheme, the U-shape design can be written $(a,b)(a,b)$ as pictured by the cyan ray in Fig.\ \ref{f:raytracing}(a), as the marginal ray that reflects from point $a$ on the first OAPM then reflects from $b$, $a$ and $b$ points on the subsequent mirrors. 
The step-shape can be written $(a,b)(b,a)$ and the S-shape is $(a,a)(a,a)$.
Retaining the brackets is useful as a guide to show where the focii of the optical system are, namely before and after each bracket.

To assess the performance of each setup, we introduced an offset in the position of the point source in the emitter (object) plane, either in the $y$-direction, as illustrated in Fig.\ \ref{f:OAPM}, in the $x$ direction, or in both. 
As a consequence of the point source being off-axis by a distance $\delta=2$\,mm, rays propagate at a slight angle 1.5$^\circ$ after the first OAPM, in comparison to the on-axis rays. 
This tilt can be observed in Fig.\ \ref{f:raytracing}(d), where the off-axis rays are represented in green.
The second OAPM, which forms an $(a,b)$ orientation pair, focused rays to a point at a distance $-\delta$ in $y$ from the optical axis, where the negative sign signifies that the image formed after the two OAPMs was inverted, and at a distance along the $z$ axis after from the SP.
In the sagittal plane (not shown here) the focus forms before the SP.
Notably, for the alternative orientation of the second OAPM, $(a,a)$, pictured in Fig.\ \ref{f:raytracing}(e), the off-axis rays are refocussed at the SP.

Geometric optics provides a rudimentary understanding of aberrations, but it has limitations as it does not consider diffraction or interference effects. 
In contrast, the physical optics module of ZEMAX can rigorously model the coherent propagation of light through the OAPMs, taking into account diffraction and interference. 
The beam is numerically defined as an array of sampled points with a complex amplitude in the plane normal to the chief ray. 
Here we performed physical optics calculations at 1\,THz, with a Gaussian initial beam profile in the emitter plane, with beam waist $w_{0} = 0.4$\,mm in $x$ and $y$ directions, initial power 1\,W and hence peak irradiance $I_0=4$\,Wmm$^{-2}$. 
While the power and irradiance here were arbitrary, the beam waist is typical of some pulsed THz sources, such as large area photoconductive emitters \cite{Mosley2017} and optical rectification from amplified laser pulses.
With this initial beam waist the beam diverges to fill the first OAPM without substantial loss at the edges, owing to the mirror's finite diameter.
In order to achieve the best numerical accuracy, the physical optics module automatically chooses between propagating light from surface to surface along $z$ via different algorithms (either Fresnel diffraction, or an angular spectrum propagation approach).
Further, it adjusts the spatial grid to ensure the correct sampling as beams change diameter.
Finally, the physical optics approach uses Gaussian ``pilot'' beams, propagated through the optical system, in order to aid the numerical algorithms: here separate $x$ and $y$ pilot beams were enabled to help better match the asymmetric shape of the OAPMs.

\section{Results}

To better investigate the impact of OAPM orientation on the capabilities of THz systems, we now consider the spatial and temporal performance of the different designs in more detail, using both ray optics and physical optics calculations.
To comprehensively assess image quality, we conducted a comparative analysis of the spatial distribution and the temporal evolution of the THz field across all geometries, at a frequency of 1\,THz (wavelength $\lambda=300\mu$m).
Note that the only impact of the wavelength is that it changes the diameter of the diffraction-limited spot size.
We set the THz beams to radiate from different positions in the emitter plane, in order to assess the impact of transverse mis-alignments and of imaging quality.
Throughout this work we present results with an offset $\delta=2$\,mm as representative of the typical level of misalignment in coarsely-positioned THz setups, and also as a size representative of the mm scale over which far-field THz imaging can be performed.
It should be noted that while results obtained with $\delta=2$\,mm are informative, larger $\delta$ produces more significant aberrations.

\subsection{Spatial performance}
\begin{figure}[!t]
	\centering
	\includegraphics[width=0.95\columnwidth]{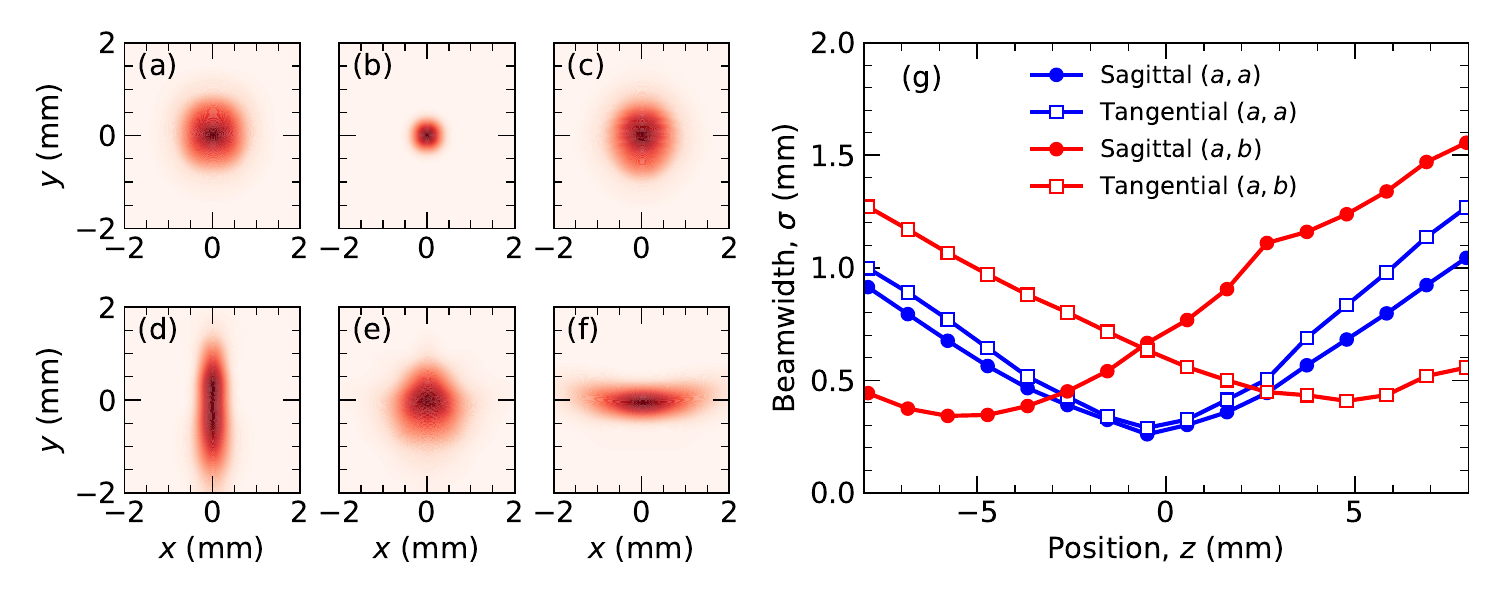}
	\caption{(a)-(c) Beam profiles of $(a,a)$ geometry from physical optics calculations for a $(0,2$\,mm$)$ offset source at the emitter, at different $z$. (a) $z=-5$\,mm relative to the focus formed (at $f=76.2$\,mm from the second OAPM), (b) at $z=0$ (the focus) and (c) $z=5$\,mm. Note that the $x$ and $y$ co-ordinates are shown relative to the beam centres (the beam is actually centred at $y=-2$\,mm relative to the optical axis). 
 (d)-(f) are the same as (a)-(c), but for positions close to the focus formed for the $(a,b)$ geometry.
 (g) Illustrates the beamwidths $\sigma$ in $x$ (sagittal, filled circles) and $y$ (tangential, open squares) calculated from the second moment of the profiles at different positions $z$ along the beam (relative to the SP, at $z=0$), for both $(a,a)$ (blue lines) and $(a,b)$ (red) oriented OAPMs. }
	\label{f:beamProfiles}       
\end{figure}

The physical optics models produced beams that propagated through the spectrometers with close to Gaussian profiles near the focii. 
Pictured in Figure \ref{f:beamProfiles}(a)-(c) are the beam profiles (normalised irradiance maps relative to the beam centres) at different positions along the propagation direction near the SP focus for the $(a,a)$ orientation, with a THz beam radiating from $(0,2$\,mm$)$ in the emitter plane.
A circular beam profile was obtained before (panel a) and after (c) the focus, with a small spot in the focal plane (panel b).
In contrast, before and after the focus the $(a,b)$ geometry results in a highly asymmetric beam profile (panels d and f) and a larger spot size in the focal plane than for the $(a,a)$ design.
These results from numerical physical optics reinforce the conclusions drawn from Gaussian beam theory \cite{Murphy1987} and ray tracing \cite{Bruckner2010} that the 4$f$ $(a,a)$ geometry can cancel out the geometric aberrations of each OAPM, and obtain good imaging performance even for off-axis THz beams.

The aberrations are not cancelled out in the $(a,b)$ orientation, with the consequence that the beam becomes asymmetric and exhibits an astigmatic difference $d$: the beam is narrowest in $x$ and $y$ at different positions along $z$, separated by a distance $d$.
This can be further seen in Fig.\ \ref{f:beamProfiles}(f), where the beamwidths in $x$ and $y$ were calculated using the second order moment of the irradiance according to the ISO 11146-1 standard. 
For the $(a,a)$ geometry the sagittal ($x$, blue circles) and tangential ($y$, blue squares) beamwidths were similar, and were smallest (around 0.3\,mm) at the SP ($z=0$). 
The sagittal (red circles) and tangential (red squares) beamwidths for the $(a,b)$ geometry had minima separated by a distance $d=10$\,mm.
The astigmatic difference, $d$, grows for larger THz beam offsets $\delta$ in the emitter plane.

We used ray tracing and spot diagrams to assess the image quality for a wide variety of initial beam offsets, as ray optics calculations are substantially faster than physical optics simulations, and capture many of the same features. 
In ray tracing, whether or not an optical system can achieve diffraction-limited performance can be assessed by examining ``spot diagrams,'' which show the spatial distribution of rays as they intersect the focal plane, and comparing it to the diffraction-limited beam size, for instance, given by an Airy disk or from physical optics. 
Alternatively, because ray tracing keeps track of the phase of the electric field, the optical path difference (OPD) of different rays can be determined relative to the chief ray.
The system can be assumed to be diffraction-limited if the phase change for all rays is sufficiently small: an OPD of less than a quarter wavelength is a common convention used in the optical modelling community to define diffraction-limited performance.

\begin{figure}[bt]
	\centering
	\includegraphics[width=0.95\columnwidth]{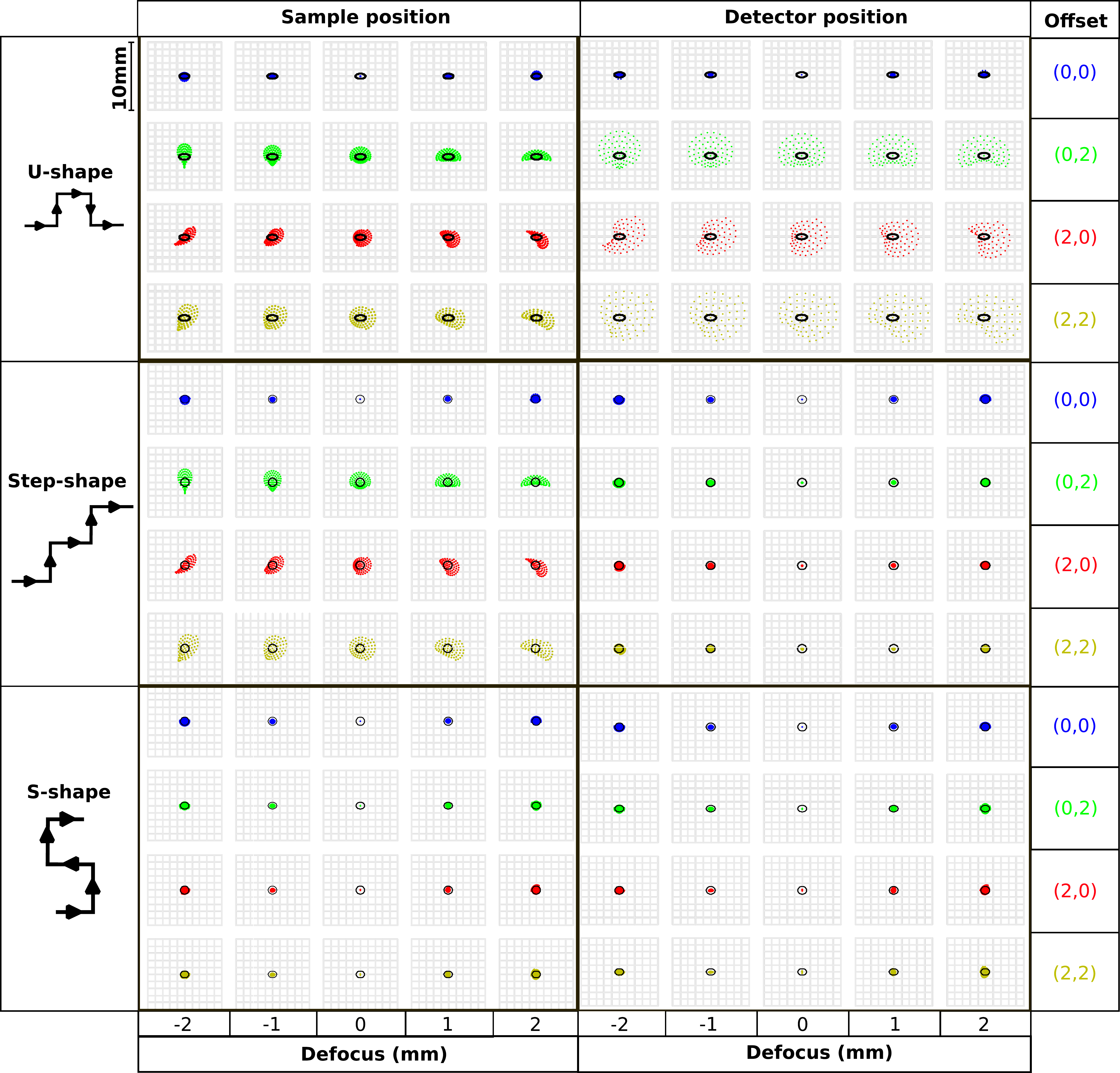}
	\caption{Spatial profile of the THz beam near the sample plane (left column) and detector plane (right column), for the three different designs (rows). Within each black rectangle, spot diagrams are shown for a particular plane and design, as a function of the offset of the source in the emitter plane and the ``defocus'' - the difference along $z$ from the ideal focus. The colours and far right-hand labels indicates the $(x,y)$ offsets of the point source in the emitter plane (in mm). Every plot has an Airy-disk (black line) that provides a visual indicator of the diffraction-limited beam size at 1\,THz ($\lambda=300\mu$m) for an entrance pupil diameter of 50.8\,mm. The U-geometry has significant aberrations for a $2$\,mm offset, both at the SP and at the DP: rays fall outside the Airy disk limit, and the image is distorted. The DP performance of the step-geometry is poor at the SP but good at the DP, while the S-shape rays fall within the Airy disk for both SP and DP, and it exhibits the least aberrations.}
	\label{f:spotdiagram}       
\end{figure}

We report spot diagrams in Fig.\ \ref{f:spotdiagram} for each geometry (three rows) and at both the sample plane (left-hand column) and in the detector plane (right-hand column). 
In each case, the sequence of spot diagrams shows how the beam propagates through the focus in steps of 1\,mm along the beam propagation direction. 
Different colours represent the spot diagrams obtained for on-axis and off-axis point sources. 
The spot diagrams are made up of a large number of spots, not all of which are separable on the scale shown here (10\,mm by 10\,mm for each spot diagram), where 1\,mm corresponds to one of the smaller grey squares.
It is evident for all geometries that propagation from an on-axis point source (blue points) is well-behaved, in that the rays traverse the system and reach a point-like focus both in the SP and the DP.

Considering the spot diagrams for the SP for the U-shape and the step-shape (top-left and middle-left black rectangles), which are identical owing to sharing the same initial $(a,b)$ geometry, it is clear that a small offset of 2\,mm in $x$ or $y$ is sufficient to substantially distort the shape of the beam through the focus in a way that is entirely consistent with the physical optics results. 
The offset in $y$ (green) creates a beam that is extended along $y$ at a defocus of -2\,mm (before the focus), and stretched along $x$ after the focus.
The rays cover a range about 2\,times wider than the diffraction-limited size at 1\,THz, as evident from a comparison with the black circles (Airy disks), which have radius $\phi_r = 1.22 \lambda/f_{\#}$ for f-number $f_{\#}$.
The system does not have diffraction-limited performance at 1\,THz or higher frequencies, as the Airy disk reduces in diameter at higher frequencies.

In contrast, for the S-shaped design, with $(a,a)$ OAPMs, the performance at the sample position (bottom-left) is excellent for all the off-axis points pictured.
The reason for the better performance can be intuitively understood as follows: at the second OAPM the marginal rays hit a matched surface, such that the distortion in the wavefront introduced by the first OAPM can be ``inverted'' by the second OAPM, as the beam propagates to the focus.
Considering the $(a,a)$ configuration further, we found it continued to achieve diffraction-limited performance for offsets as large as $\pm4$\,mm, although for even larger offsets (\emph{e.g.}\ 8\,mm) the beam tilt in the tangential plane in the collimated section became too substantial and a large fraction of marginal and central rays were lost.
This limits the maximum area for useful THz imaging (using these OAPM diameters) to around $\pm5$\,mm from the optical axis.

We continue to discuss the spot diagrams in Fig.\ \ref{f:spotdiagram} by considering the right-hand column, which illustrates the spot diagrams close to the detector plane for the same point source offsets and defocus positions as discussed above. 
From these diagrams, it is clear that the U-shape, $(a,b)(a,b)$, and step-shape, $(a,b)(b,a)$, designs have substantially different imaging performances at the detector plane.
For the U-shape, the aberrations evident at the SP are further compounded by the second OAPM pair, such that the off-axis spots are even larger in the DP.
In contrast, for the step-shape geometry, the DP performance is diffraction-limited and comparable to that of the S-shape.

\begin{figure}[tb]
	\centering
	\includegraphics[width=0.8\columnwidth]{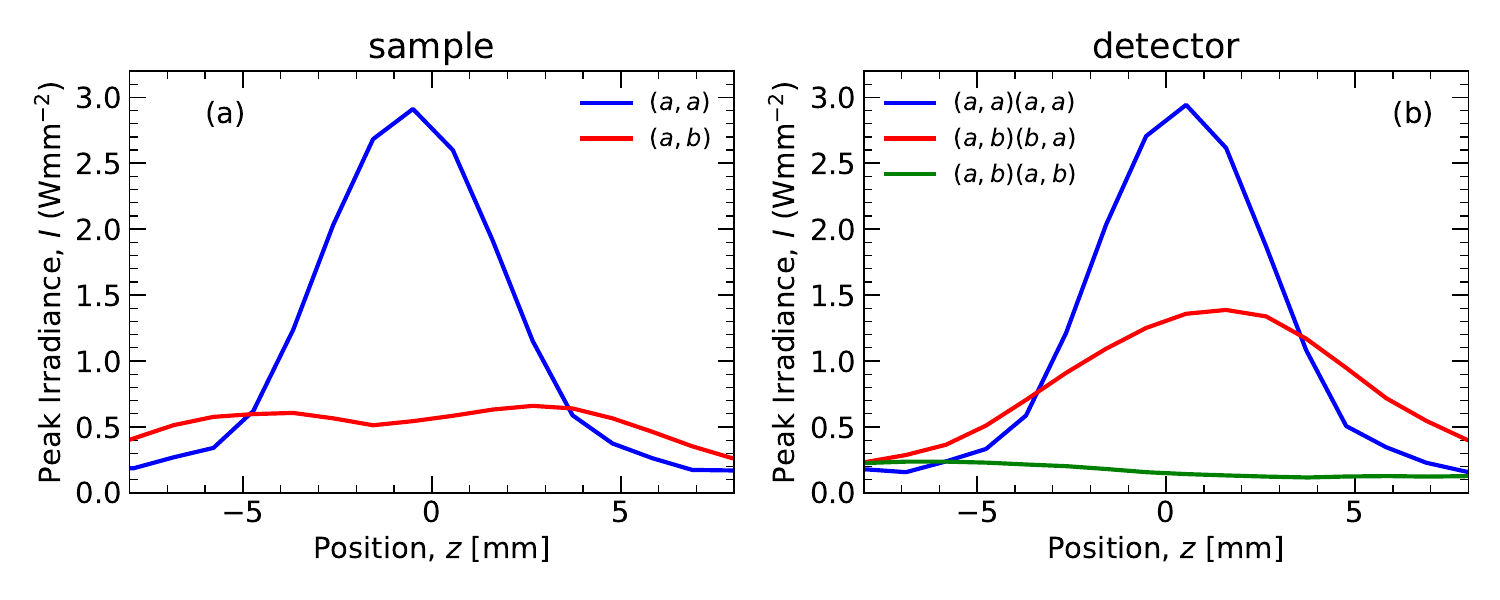}
	\caption{Peak irradiance extracted from physical optics calculations for an initial Gaussian beam centred at $(0,2$\,mm$)$ in the emitter plane, shown (a) near the sample focal plane and (b) near the detector focal plane.}
	\label{f:irradiance}       
\end{figure}

Returning to the diffraction-based results, these allow the irradiance of the beam to be readily calculated, as reported in Fig.\ \ref{f:irradiance} for gaussian beams originating from $(0,2$\,mm$)$ in the emitter plane, away from the optical axis.
Near the sample position, shown in Fig.\ \ref{f:irradiance}(a), two maxima are formed before and after the nominal focal length (shown as $z=0$) as a result of the astigmatic difference.
These correspond to positions close to the minimum beam waists in $x$ and $y$ (Fig.\ \ref{f:beamProfiles}(b)).
In contrast, the $(a,a)$ geometry has a smaller beam waist and negligible astigmatic difference, resulting in a 5 times greater peak irradiance for the $(a,a)$ geometry than the $(a,b)$ geometry.
Similarly to the spot diagrams near the detector plane (Fig.\ \ref{f:spotdiagram}), the irradiance near the detector plane (Fig.\ \ref{f:irradiance}(b)) demonstrates that the $(a,a)(a,a)$ and $(a,b)(b,a)$ designs perform far better than the $(a,b)(a,b)$ geometry.

\subsection{Temporal performance}

\begin{figure}[tb]
	\centering
	\includegraphics[width=1.0\textwidth]{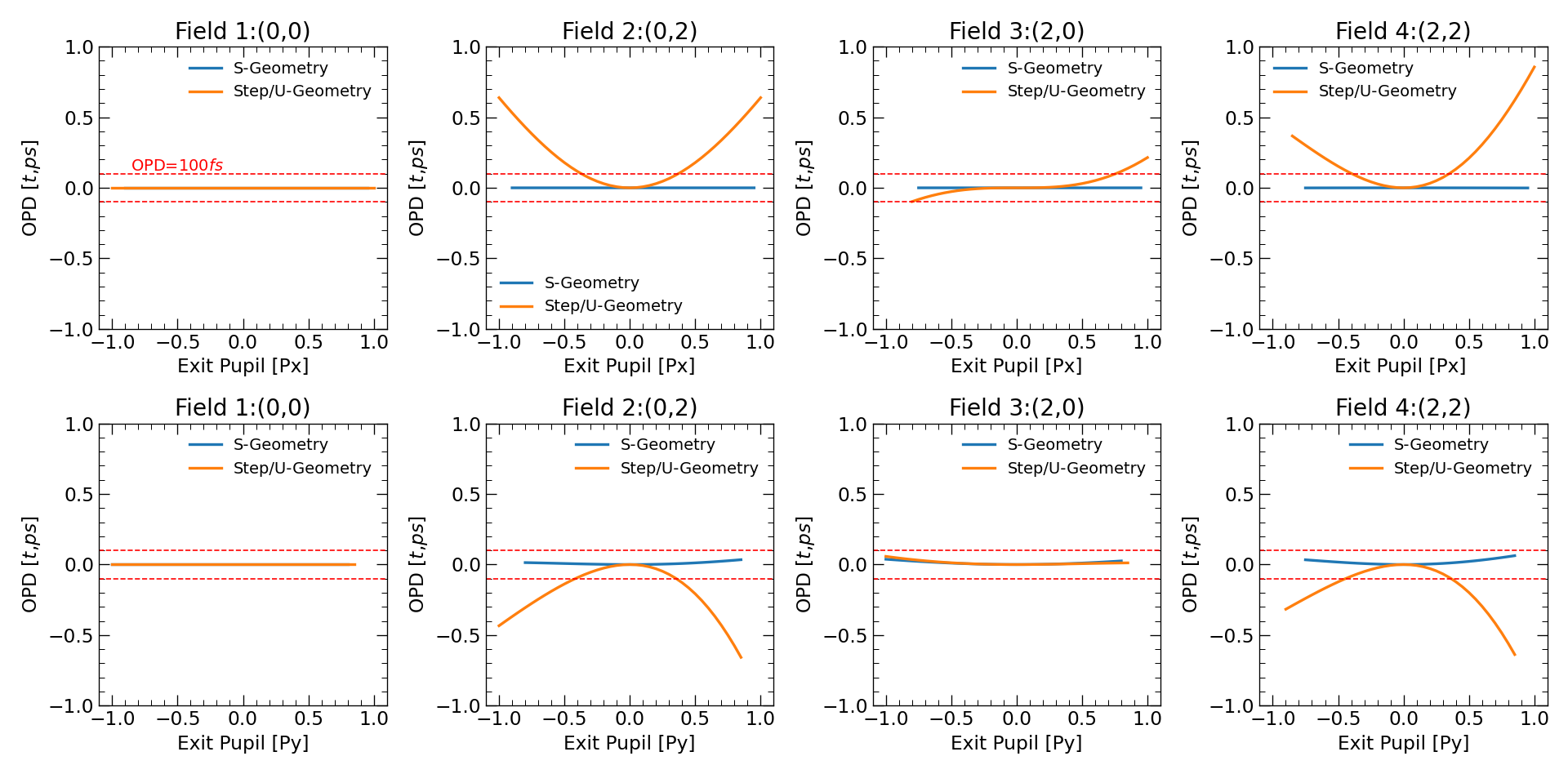}
	\caption{Wavefront distortion illustrated by the optical path difference for the first two OAPMs. The plot presents the tangential ($P_y$) and sagittal plane ($P_x$) optical path differences at the exit pupil. The dashed red line denotes the typical fs laser pulse duration (100\,fs) used in THz-TDS setups. The OPD is not defined for exit pupil coordinates where there are no rays on the final OAPM (because of beam tilt), for example along $P_y$ in the S-geometry for $(0,2$\,mm$)$.}
	\label{f:OPDsample}
\end{figure}

Spot diagrams (from ray optic calculations) and intensity profiles (from physical optics simulations) offer a purely geometric perspective on aberrations, and provide insights into their spatial characteristics. 
However, for a more comprehensive understanding, it is advantageous to investigate the optical path difference (OPD) of the system relative to the ideal wavefront surface, using ray optics. 
The ideal wavefront is taken to be a sphere with radius $f$, centred at the chief ray's position in the image plane. 
The OPD is then calculated for the different rays and is shown graphically by plotting it as a function of positions $P_x$ and $P_y$, which correspond to the ray's position in the exit pupil (the last optic) in the sagittal and tangential planes, respectively. 
By convention $P_x$ and $P_y$ are given normalised by the radius of the last optic.
In Figure \ref{f:OPDsample} we present the OPD for the sample plane versus $P_x$ (top row) and $P_y$ (bottom row), for various point source positions in the emitter plane (left to right), and for each design.
While the OPD is often shown as the path difference divided by the wavelength, we opted here to show the OPD as a time delay to allow a better understanding of the scale of the OPD changes for the THz time-domain spectroscopy community.

For the on-axis case, $(0,0)$, the horizontal lines in the top-left and bottom-left panels in Fig.\ \ref{f:OPDsample} signify the absence of an optical path difference (OPD) for all designs, i.e.\ zero phase difference, representing a perfect spherical wavefront converging to a point image. 
The off-axis points (representing translational misalignment or finite beam size) result in the incident wavefront converging at the focal point with a slight tilt caused by either the delayed or advanced arrival of these rays. 
For the $(0,2$\,mm$)$ field point a parabola-like OPD can be seen in $P_x$ and $P_y$ for the step and U geometries, with large OPD values $>500$\,fs for the marginal rays (rays at large positive or negative values of $P_x$ or $P_y$). 
In contrast, for S-geometry, the OPD change is small.

For THz time-domain spectroscopy applications, it is important to assess how the different path lengths impact the THz pulse duration at the sample position.
One could average over the OPD curve by using an assumed profile for the THz beam at the exit pupil (\emph{e.g.}\ a Gaussian) to weight the amplitude for each ray.
Rather than perform that exercise, which would be wavelength and beam specific, we instead highlight the typical pulse duration of the fs laser pulses ($\sim100$\,fs) used in THz time-domain spectroscopy by the horizontal dashed lines.
An OPD larger than this limit will introduce a substantial phase shift, and hence will broaden the duration of a THz pulse, simultaneously lowering the peak amplitude of the THz electric field.
It is clear from Fig.\ \ref{f:OPDsample} that for off-axis points (e.g.\ from poor emitter alignment) the temporal duration of the THz pulse at the sample plane is likely to be longer in duration for the step- and U-geometries, as a result of geometric aberrations.
Further, it can be seen that precise alignment in the $y$ direction is more critical (the OPDs are smaller for offsets in $x$).

\begin{figure} [tb]
	\centering
	\includegraphics[width=1.0\textwidth]{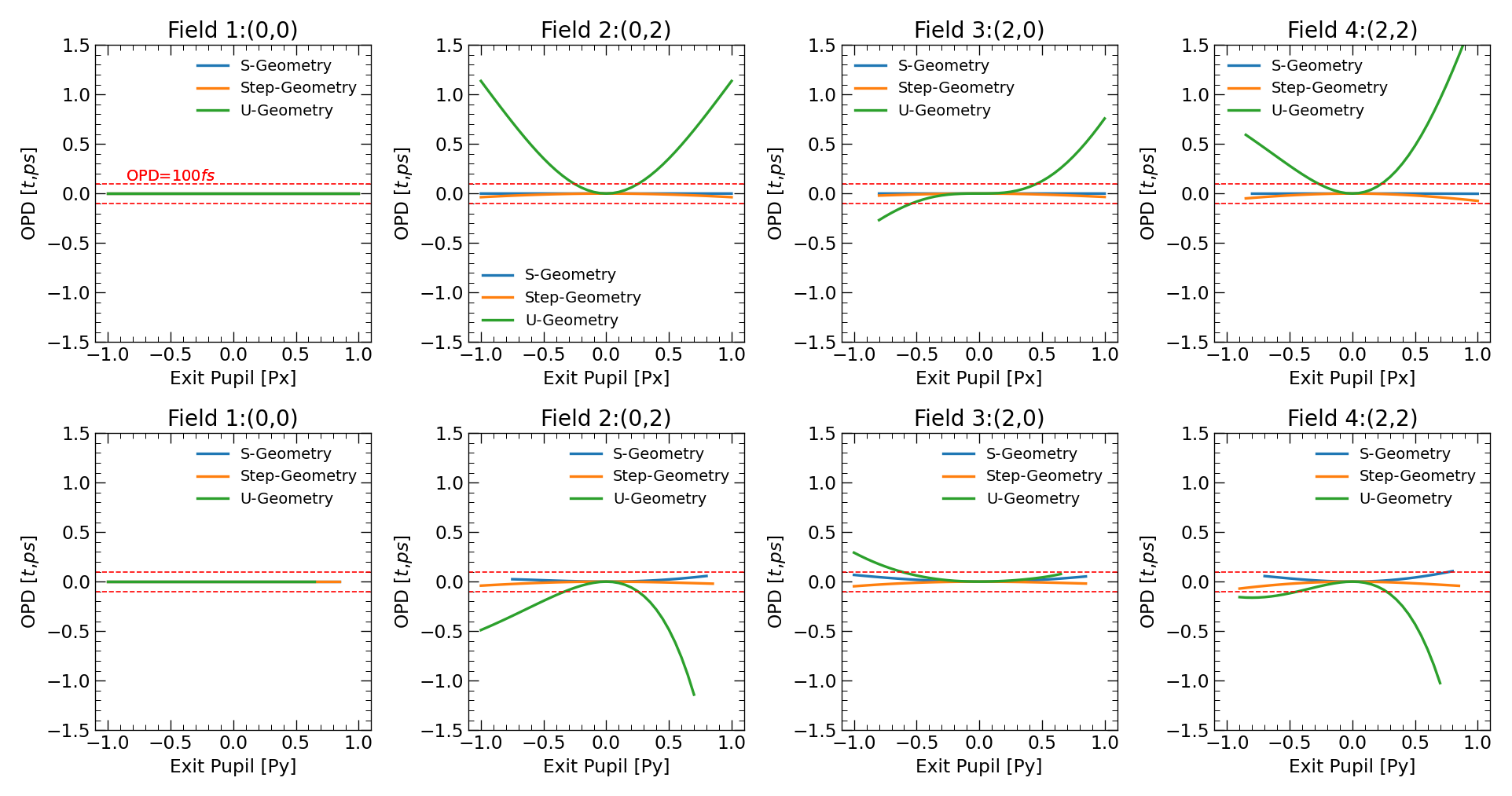}
	\caption{Wavefront distortion illustrated by the optical path difference for the entire optical path to the detection plane. The S- and step-geometry (blue and orange respectively) demonstrate a superior performance for off-axis points, showing minimal path difference and the least temporal broadening. On the other hand, for the U-geometry, we observe that translational misalignment rapidly cause wavefront aberrations leading to large OPDs.}
	\label{f:OPDdetector}       
\end{figure}

The OPD for the complete 4 OAPM optical setups is illustrated in Fig.\ \ref{f:OPDdetector}, which provides a clear depiction of why the aberrations in the DP differ for the three geometries.
The OPD for off-axis point sources is largest for the U-geometry, reaching as large as 1\,ps (green curves), suggesting that the U-shape will temporally broaden the THz pulse at the detection position as a result of the substantial geometric aberrations.
In comparison, the step-shape (orange curves) and S-geometry (blue curves) have a phase difference near zero, and within the $\pm100$\,fs limit that we suggest is important for typical THz time-domain spectrometer systems. 

While the excellent performance of the S-geometry can be assumed to result from the cancellation of distortions within each pair in the $(a,a)(a,a)$ orientation, the similar performance of the step-shape at the detector position, and the poor performance of the U-shape need further explanation.
The relative performance can be explained by considering the OPD introduced by each OAPM pair, as well as keeping track of the relative orientation of each pair using the marginal ray notation introduced above: $(a,b)(b,a)$ for the step-shape and $(a,b)(a,b)$ for the U-shape.
For the first OAPM pair, the $(a,b)$ orientation creates a distorted wavefront for an off-axis field point, for example, the roughly parabolic OPDs in $P_x$ and $P_y$ seen in Fig.\ \ref{f:OPDsample} for a $(0,2$\,mm$)$ point source in the emitter plane.
In the step-shape, the second OAPM pair has $(b,a)$ orientation, and is in the same orientation in the lab coordinate system as the first pair.
However, the OPD that the $(b,a)$ pair produces is not the same as for the first $(a,b)$ pair, because the image is inverted from the EP to the SP: e.g.\ the $(0,2$\,mm$)$ point in the emitter plane maps to $(0,-2)$\,mm in the sample plane.
Critically, the OPD flips in sign for inverted field points in the $x-y$ plane \cite{Bruckner2010}. 
Thus the OPD (or wavefront distortion) created by the second, $(b,a)$ pair in the step-geometry acts to cancel out the OPD of the first pair.
In contrast, if the second pair has $(a,b)$ orientation, as in the U-shape, the OPD of the second pair has the same sign, and the wavefront distortions add rather than subtract.

\section{Discussion}

From the results presented above we can conclude that the S-shape geometry should be adopted in linear, non-linear and OPTP spectroscopy applications, as it is the most robust to the misalignment of the THz source in the $x-y$ plane, and can best cancel out aberrations introduced by the OAPMs in order to achieve a diffraction-limited performance and the highest irradiance in both the SP and DP.
For linear THz spectroscopy, how critical diffraction-limited performance is at the intermediate focal plane (the sample position) depends on the application: if raster scanned images are required, or if samples with small $x-y$ extent are studied, then it is clearly advantageous to also work with the S-shape $(a,a)(a,a)$ geometry.
However, if only spectroscopic measurements are desired, and the highest peak electric field/smallest beam area at the sample position is not critical, then the step-shape can be alternatively used as it has reasonable performance in the DP.

While the magnitude of the geometric aberrations in the SP and DP are properties of the spectrometer design and the EP offset, whether or not diffraction-limited performance can be achieved depends on the wavelength and it is therefore useful to consider the frequency-dependent performance of each design. 
For example, diffraction-limited performance is not achieved at 1\,THz in the SP for the U- and step-geometries (Fig.\ \ref{f:spotdiagram}) for a 2\,mm offset, because geometric aberrations cause rays to fall over a region about twice the diameter of the Airy disk.
However at 500\,GHz, where the Airy disk doubles in radius relative to that at 1\,THz, the majority of rays would fall within the Airy disk and diffraction-limited performance would be achieved at the SP.
For higher THz frequencies, the Airy disk reduces in diameter and the relative contribution of aberrations becomes more significant.

\begin{figure} [tb]
	\centering
	\includegraphics[width=1.0\textwidth]{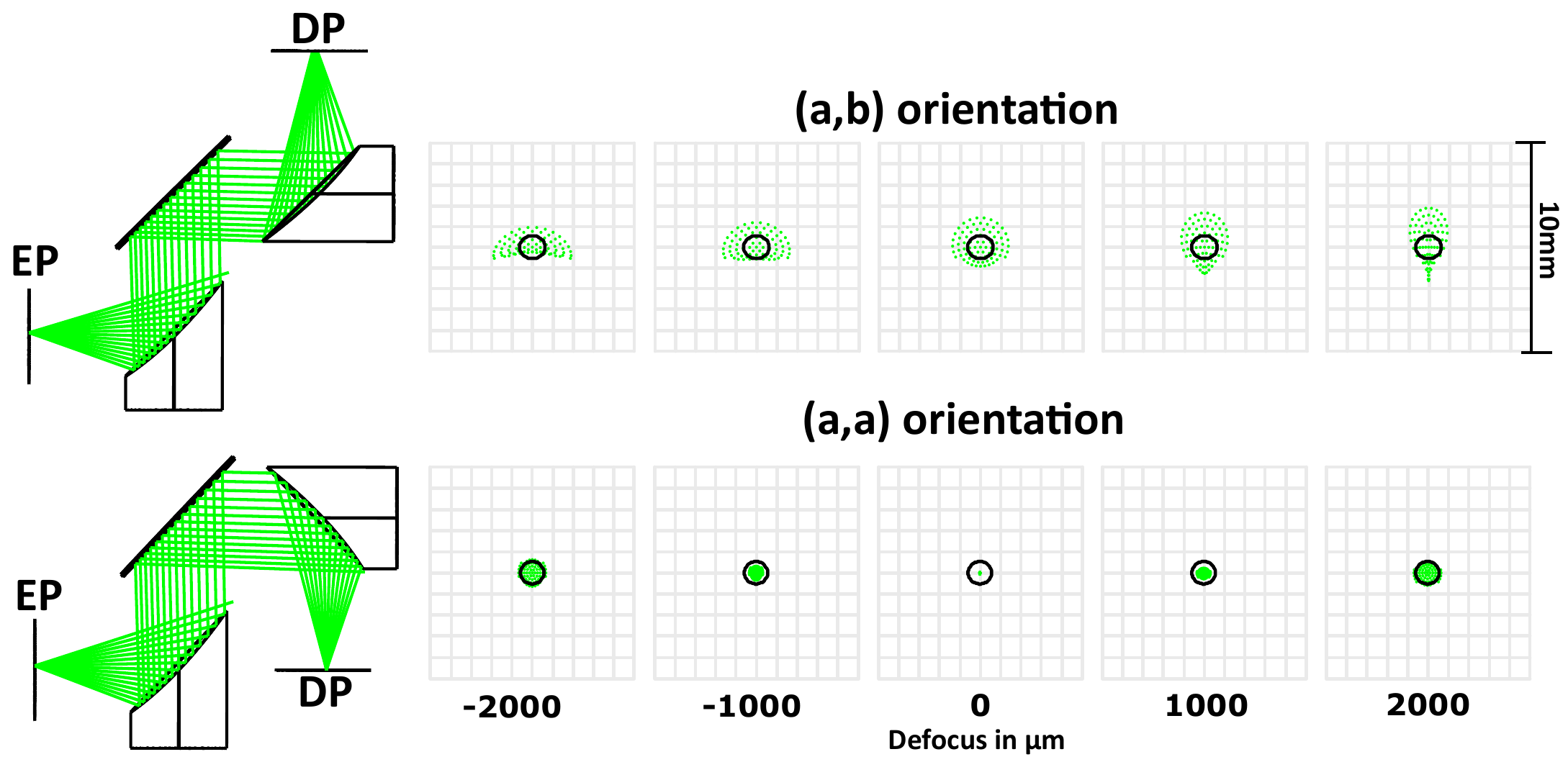}
	\caption{Two OAPM geometry including a folding mirror, for a $(0,2$\,mm$)$ point source. Top row: $(a,b)$ orientation (left) and through-focus spot diagrams (right). Bottom row: as first row, but for the $(a,a)$ orientation of the second OAPM.}
	\label{f:reflection}       
\end{figure}

Based on a consideration of the spot diagrams and OPDs presented above, we can now suggest rules to aid the design of OAPM systems for THz spectroscopy and imaging:
\begin{enumerate}
    \item geometric aberrations may be expected to impact the optical performance of typical OAPM systems at 1\,THz and higher frequencies;
    \item an OAPM pair should have $(a,a)$ orientation to minimise aberrations at the focus;
    \item if an $(a,b)$ orientation is imposed by some other experimental constraint, then it will have aberrations in its image plane;
    \item these aberrations can be inverted by a $(b,a)$ pair.
\end{enumerate}
These rules allow the optimum OAPM orientation to be deduced for more complex spectrometer designs, without needing to ray-trace the geometry. 
For example, if an additional optic such as a mirror is required to change the THz beam direction, or to couple an optical beam along the optical axes of the OAPMs (using a THz mirror that passes visible light and reflects THz radiation), then it is desirable to know the optimum orientation for the OAPM(s) after the THz mirror.
We now consider the setup pictured in Fig.\ \ref{f:reflection}, where a mirror is placed after the first OAPM, and the THz beam propagates along $z$. 
To form a focus, two orientations of the second OAPM are possible (top and bottom left diagrams).
To identify the correct orientation of the final OAPM, such that aberrations are minimised, marginal ray notation and the rules above are useful.
According to these rules the $(a,a)$ configuration (bottom left) should be used as it will have less significant geometric aberrations than the $(a,b)$ configuration (top left). 
Indeed the spot diagrams in the DP (right hand panels) validate this assertion.
Similar considerations can be applied to correctly orient OAPMs in reflection geometry systems, where the THz beam reflected from a sample is collected, and THz beamsplitters are used.

For completeness we briefly discuss the THz detection process.
In THz detection with photoconductive antennae, the typical antenna gap is around 10-20\,$\mu$m, which is strongly sub-wavelength. 
The effective width of the antenna is larger than its gap, and varies with geometry (e.g.\ bow-tie, dipole, log-spiral). 
Hence commercial photoconductive antennae use high numerical aperture silicon lenses to focus THz beams down to sizes smaller than the free-space limit. 
Here the numerical model does not treat metallic antennae or silicon lenses, and our numerical results are hence closest to the experimental case of electro-optic sampling, where no lens or antenna is used. 
More advanced electromagnetism calculations would be required to include antenna effects. 
However, the conclusions drawn here are independent of the detection process: for example the electric field input into the detector will be less distorted for the $(a,a)(a,a)$ geometry than the $(a,b)(a,b)$ geometry.

Finally, it is worth discussing the impact of geometric aberrations on polarisation. 
There have been studies that reported asymmetric field and polarization distributions for OAPMs with different f-numbers used in the U- and step-shape \cite{Sung}. 
Conversely, abrupt changes in THz polarization states were experimentally observed by Takai \emph{et al.} \cite{takai2016spatial}, and were ascribed to the inherent geometry of the OAPMs. 
In the present study, the polarization performance of different OAPM geometries was not considered. 
We note that in a recent experimental study using multi-pixel THz emitters producing radial and azimuthal polarization, the polarization state can be experimentally corrected \cite{Deveikis2022}.

\section{Conclusion}

The study presented here contributes to a better understanding of the propagation and behaviour of THz pulses in multi-OAPM geometries. 
We demonstrated that geometric aberrations can limit the optical performance both in the sample and detector plane: they increase the spot size and decrease the amplitude of the THz field (they lower the irradiance).
Depending on the specific orientation of each OAPM in the optical system, the wavefront distortions, and aberrations at each subsequent focus, can be reduced or enhanced.
Results from both ray tracing and physical optics, including diffraction and interference effects, were in agreement.
We introduced marginal ray notation to capture the orientation of each OAPM in a system.
The S-shape cancels out the geometric aberrations by the correct orientation of the second OAPM for each pair throughout the optical path, while the step-shape achieves good performance for off-axis rays at the detection plane, by the second pair of OAPMs cancelling out the distortion produced by the first pair. 
The use of the U-shape design should be discouraged as it adds, rather than subtracts, the distortion from each OAPM pair.
Our modelling approach and design rules can be readily applied to the design of more complex THz imaging and spectroscopy setups based on OAPMs.

\section*{Declarations}
\textbf{Ethical Approval.} N/A
 
\noindent\textbf{Competing interests.} The authors declare that they have no competing interests.
 
\noindent\textbf{Authors' contributions.} N.C. performed the optical modelling, prepared the figures, and drafted and edited the paper. J.L. discussed the optical modelling, helped prepare the figures, and drafted and edited the paper.
 
\noindent\textbf{Funding.} The authors would like to acknowledge funding from the EPSRC (UK) (Grant No.\ EP/V047914/1).
 
\noindent\textbf{Availability of data and materials.} ZEMAX modelling files and data are available from the authors on reasonable request.

\bibliography{references}
\bibliographystyle{unsrt}


\end{document}